\documentclass[epj,nopacs]{svjour}
\usepackage{latexsym}
\usepackage{graphics}
\usepackage{amssymb}
\usepackage{amsmath}

\begin{document}
\thispagestyle{empty}
\title{Exact Casimir-Polder potential between
 a particle and an ideal metal cylindrical shell and
 the proximity force approximation}

\author{
V.B.~Bezerra\inst{1}, E.R.~Bezerra~de~Mello\inst{1},
G.L.~Klimchitskaya\inst{1,2},
 V.M.~Mostepanenko\inst{1,3,}\thanks{E-mail:
Vladimir.Mostepanenko@itp.uni-leipzig.de},
A.A.~Saharian\inst{4,}\thanks{E-mail:
aram.saharian@gmail.com} }

\institute{
Department of Physics, Federal University of Para\'{\i}ba,
C.P.\ 5008, CEP 58059--900, Jo\~{a}o Pessoa, Pb-Brazil
\and
{North-West Technical University,
Millionnaya Street 5, St.Petersburg,
191065, Russia}
\and
{Noncommercial Partnership ``Scientific Instruments'',
Tverskaya Street 11, Moscow,
103905, Russia}
\and
{Department of Physics, Yerevan State University, 1 Alex
Manoogian Street, 0025, Yerevan, Armenia} }

\date{Received: date / Revised version: date}

\abstract{
We derive the exact Casimir-Polder potential for a polarizable
microparticle inside an
ideal metal cylindrical shell using the Green function method.
The exact Casimir-Polder potential for a particle  outside a
shell, obtained recently by using the Hamiltonian approach,
is rederived and confirmed.
The exact quantum field theoretical
result is compared with that obtained using the proximity force
approximation and a very good agreement is
demonstrated at separations below 0.1$R$, where $R$
is the radius of the cylinder. The developed methods
are applicable in the theory of topological defects. }

\authorrunning{ V.~B.~Bezerra et al.}
\titlerunning{The exact Casimir-Polder potential
 and the proximity force approximation}

\maketitle

\section{Introduction}

The Casimir-Polder and Casimir forces act between a polarizable
microparticle and a surface and
between two surfaces, respectively \cite{1,2}. Both forces are of quantum
nature and originate from zero-point oscillations of quantized fields. The
Casimir force due to nontrivial spatial topology plays an important role in
the compactification of extra dimensions in Kaluza-Klein theories \cite{2a},
in cosmology \cite{2b}, and in the theory of topological defects \cite{2c}. With
advances in nanotechnology and successes in the production of ultracold
atoms it became possible to measure the Casimir and Casimir-Polder force
with increased precision (see \cite{3,4} for a review).
Interestingly, some of the fundamental results on the Casimir effect
have already found topical applications. As an example,
the Casimir energy for isolated cylindrical and toroidal carbon
nanotubes was calculated in \cite{12a,12b}
basing on the theory of topological defects.

Many calculations of the Casimir and other interactions between
curved surfaces were performed
by means of
the so-called \textit{proximity force approximation} (PFA) \cite{15,15a,15b,22a}.
This is an approximate method which replaces infinitesimal elements of
curved surfaces with pairs of plane parallel plates and uses advantages of a
known exact solution for the plane parallel geometry. The PFA is commonly
used for comparison between experimental results and theory \cite{21a}.
This
method is  subject to some restrictions and possesses only a limited
accuracy. Because of this, a more fundamental quantum field theoretical
approach is highly desirable.
Such an approach was developed, for instance, for an ideal metal
cylinder above an ideal metal plane \cite{18N}.
For this configuration, the asymptotic
expansion of the exact result in the limit of short separations
was found and compared with the PFA \cite{19N}.
As one more configuration allowing
 the fundamental description
based on first principles of quantum field theory one could consider the
interaction potential of a
microparticle with an ideal metal cylindrical shell. This
configuration has long been studied (see related references in \cite{21b}).
Recently, the exact interaction potential for a microparticle external to a
cylindrical shell has been found in \cite{21b} using the Hamiltonian
approach.
It was also suggested \cite{add} to use
Rydberg atoms out of thermal equilibrium inside a cylindrical cavity
for the observation of resonant Casimir-Polder interactions.

Here, we derive the exact expression for the Casimir-Polder potential
between a polarizable microparticle either external or internal to  an ideal metal
cylindrical shell using the Green tensor of the electromagnetic field
in a
cylindrical configuration. For a microparticle external to a cylindrical shell the
obtained results are shown to be in agreement with those derived in \cite%
{21b} with the help of another method. We compare the exact particle-cylinder
potential with the approximate result obtained using the PFA. This allows
one to determine rigorously the application region of the PFA in one more of
rare cases in the Casimir physics where the exact
quantum field theoretical solution is available.  The
role of the dynamic polarizability of a particle is investigated. The suggested exact
approach based on the use of the Green tensor allows generalization for
different boundary conditions.

The structure of this paper is as follows. In Sect.~2 we derive the Green
tensor for an exterior region of a cylindrical shell.
 In Sect.~3 the same
is done for an interior region. Section~4 is devoted to the derivation of
the exact Casimir-Polder potential for a particle situated either in exterior
or interior regions of the cylindrical shell.
In Sect.~5 the role of dynamic
effects is considered and the comparison between the exact and the PFA
results is performed.
Section~6 contains computational results for a particle inside a
cylindrical shell. In
Sect.~7 the reader will find our conclusions and discussions.

\section{Green tensor for an exterior of ideal metal cylindrical shell}

We consider the Casimir-Polder interaction between
a microparticle (an atom or a molecule) and an ideal
metal cylinder. The corresponding interaction energy is given by the
expression (see, for instance, \cite{16})
\begin{equation}
U(\mathbf{r})=\frac{\hbar }{2\pi }\int_{0}^{\infty }d\xi \,\,\alpha
_{jl}(i\xi )G_{jl}^{\rm{(c)}}(\mathbf{r},\mathbf{r};i\xi ),  \label{UDisp}
\end{equation}%
where $\alpha _{jl}(i\xi )$ is the polarizability of a particle situated at a
point $\mathbf{r}$ calculated along the imaginary frequency axis, $j,l=1,2,3$%
, and
\begin{equation}
G_{jl}^{\rm{(c)}}(\mathbf{r},\mathbf{r}^{\prime };\omega )=\frac{1}{2\pi }%
\int_{-\infty }^{+\infty }d\tau \,G_{jl}^{\rm{(c)}}(x,x^{\prime
})e^{i\omega \tau }.  \label{Gilb}
\end{equation}%
Here, $x=(t,\mathbf{r})$, $x^{\prime }=(t^{\prime },\mathbf{r}^{\prime })$, $%
\tau =t-t^{\prime }$. In (\ref{Gilb}), $G_{jl}^{\rm{(c)}}(x,x^{\prime
})$ is the part of the retarded Green tensor $G_{jl}(x,x^{\prime })$ induced
by the cylindrical shell:%
\begin{equation}
G_{jl}(x,x^{\prime })=G_{jl}^{(0)}(x,x^{\prime })+G_{jl}^{\rm{(c)}%
}(x,x^{\prime }),  \label{Gdecomp}
\end{equation}%
where $G_{jl}^{(0)}(x,x^{\prime })$ is the Green tensor in free space. The
retarded Green tensor is defined as%
\begin{equation}
\hbar G_{jl}(x,x^{\prime })=-i\theta (\tau )\langle E_{j}(x)E_{l}(x^{\prime
})-E_{l}(x^{\prime })E_{j}(x)\rangle ,  \label{Gil}
\end{equation}%
where $\theta (x)$ is the unit-step function, $E_{i}(x)$ is the operator of
the $i$ component of an electric field, and the angular brackets mean the
vacuum expectation value.

For the evaluation of the vacuum expectation value in (\ref{Gil}) we use
the mode sum formula%
\begin{equation}
G_{jl}(x,x^{\prime })=-i\theta (\tau )\sum_{\alpha }[E_{\alpha
j}(x)E_{\alpha l}^{\ast }(x^{\prime })-E_{\alpha l}(x^{\prime })E_{\alpha
j}^{\ast }(x)],  \label{GilMode}
\end{equation}%
where $\{\mathbf{E}_{\alpha }(x),\mathbf{E}_{\alpha }^{\ast }(x)\}$ is the
complete set of eigenfunctions for the electric field in the region outside
a conducting cylindrical shell, specified by the collective index $\alpha =(\gamma
,m,k)$ (see below), and the asterisk stands for a complex conjugate. For the
geometry under consideration we have two different classes of eigenfunctions
corresponding to the cylindrical waves of the transverse magnetic (TM) and
transverse electric (TE) types. In cylindrical coordinates $(r,\phi ,z)$,
the eigenfunctions, satisfying the ideal metal boundary condition on the
surface of a cylinder of radius $R$, have the form%
\begin{equation}
\mathbf{E}_{\alpha }^{(\lambda )}=\beta _{\alpha ,\lambda }\mathbf{E}%
_{\alpha }^{(\lambda )}(r)e^{im\phi +ikz-i\omega t},  \label{Ealf}
\end{equation}%
where $m=0,\pm 1,\pm 2,\ldots $, $-\infty <k<\infty $, and $\lambda =1,2$
correspond to the TM and TE waves, respectively. For separate components of
the electric field the radial functions in (\ref{Ealf}) have the form%
\begin{eqnarray}
&
E_{\alpha ,r}^{(1)}(r) =ik\gamma g_{1,\alpha }^{\prime }(\gamma r),\quad
& E_{\alpha,r}^{(2)}(r) =-\frac{\omega m}{cr}g_{2,\alpha }(\gamma r),
\nonumber \\
&
E_{\alpha ,\phi }^{(1)}(r)=-\frac{km}{r}g_{1,\alpha }(\gamma
r),\quad
&
E_{\alpha ,\phi }^{(2)}(r)=-i\frac{\omega }{c}\gamma g_{2,\alpha }^{\prime
}(\gamma r),
\nonumber \\
&
E_{\alpha ,z}^{(1)}(r)=\gamma ^{2}g_{1,\alpha }(\gamma r),\quad
&
E_{\alpha ,z}^{(2)}(r)=0,  \label{Eirn}
\end{eqnarray}%
where here and below $\omega =c\sqrt{\gamma ^{2}+k^{2}}$, $0\leq \gamma
<\infty $, the prime means the derivative with respect to the argument, and
the following notations are introduced:%
\begin{eqnarray}
g_{1,\alpha }(\gamma r) &=&J_{|m|}(\gamma r)Y_{|m|}(\gamma R)-Y_{|m|}(\gamma
r)J_{|m|}(\gamma R),  \nonumber \\
g_{2,\alpha }(\gamma r) &=&J_{|m|}(\gamma r)Y_{|m|}^{\prime }(\gamma
R)-Y_{|m|}(\gamma r)J_{|m|}^{\prime }(\gamma R).  \label{g12}
\end{eqnarray}%
In equation (\ref{g12}), $J_{\nu }(x)$ and $Y_{\nu }(x)$ are the Bessel and
Neumann functions. The normalization coefficient in (\ref{Ealf}) is
given by the expression%
\begin{eqnarray}
&& \beta _{\alpha ,1}^{-2}=2\pi c^{-2}\gamma \omega \left[
J_{|m|}^{2}(\gamma R)+Y_{|m|}^{2}(\gamma R)\right],  \label{betalfext} \\
&& \beta _{\alpha ,2}^{-2}=2\pi c^{-2}\gamma \omega \left[ {J_{|m|}^{\prime}}%
^{\!\!\!\!\!2}\,(\gamma R)+ {Y_{|m|}^{\prime}}^{\!\!\!\!2}\,(\gamma R)\right]%
.  \nonumber
\end{eqnarray}

Substituting the eigenfunctions (\ref{Ealf}) into the mode-sum (\ref{GilMode}%
), we find for the Green tensor%
\begin{eqnarray}
&&
G_{jl}(x,x^{\prime })=-i\theta (\tau )\sum_{\alpha }\sum_{\lambda
=1,2}\,\beta _{\alpha ,\lambda }^{2}
\label{GilMod2} \\
&&~\times
\left[ T_{\alpha }E_{\alpha ,j}^{(\lambda )}(r)E_{\alpha
,l}^{(\lambda )\ast }(r^{\prime }) -T_{\alpha }^{\ast }E_{\alpha
,l}^{(\lambda )}(r^{\prime })E_{\alpha ,j}^{(\lambda )\ast }(r)\right] .
\nonumber
\end{eqnarray}%
Here, the summation over the collective index $\alpha $ means the summation
over $m$ and integrations over $k$ and $\gamma $,
and
\begin{eqnarray}
&&
T_{\alpha
}=\exp(im\Delta \phi +ik\Delta z-i\omega \tau ),
\nonumber \\
&&
\Delta\phi=\phi^{\prime}-\phi, \qquad
\Delta z=z^{\prime}-z.
\nonumber
\end{eqnarray}

In order to extract from (\ref%
{GilMod2}) the cylinder-induced part, we subtract from the right-hand side
of this equation the respective expression for the bulk without a
cylindrical shell.
The latter is given by (\ref{GilMod2}), where $\beta _{\alpha
,\lambda }^{2}$ should be replaced with $c^{2}/(2\pi \gamma \omega )$ and
the respective field components are given by (\ref{Eirn}) where the
functions $g_{\lambda ,\alpha }(\gamma r)$ are replaced with $J_{|m|}(\gamma
r)$. As a result, the cylinder-induced part of the Green tensor can be
presented in the from%
\begin{equation}
G_{jl}^{\rm{(c)}}(x,x^{\prime })=i\frac{\theta (\tau )}{4\pi }\sum_{\alpha
}\frac{c^{2}}{\gamma \omega }\sum_{s=1,2}\left[ T_{\alpha }\Xi _{\alpha
,jl}^{(s)}-T_{\alpha }^{\ast }\widetilde{\Xi }_{\alpha ,jl}^{(s)}\right] ,
\label{Gilb2}
\end{equation}%
where the following notation is used:%
\begin{eqnarray}
\Xi _{\alpha ,jl}^{(s)}&=&\frac{J_{|m|}(\gamma R)}{H_{|m|}^{(s)}(\gamma R)}%
E_{\alpha ,j}^{(1,s)}(r)\widetilde{E}_{\alpha ,l}^{(1,s)}(r^{\prime })
\nonumber \\
&&
 +\frac{J_{|m|}^{\prime }(\gamma R)}{H_{|m|}^{(s)\prime
}(\gamma R)}E_{\alpha ,j}^{(2,s)}(r)\widetilde{E}_{\alpha
,l}^{(2,s)}(r^{\prime }).  \label{Sigma}
\end{eqnarray}%
Here, the quantities $E_{\alpha ,j}^{(\lambda ,s)}(r)$ are given by (\ref%
{Eirn}), where the functions $g_{\lambda ,\alpha }(\gamma r)$ are replaced
by the Hankel functions $H_{|m|}^{(s)}(\gamma r)$. Tilde in (\ref{Gilb2})
and (\ref{Sigma}) stands for the complex conjugation which is, however, not
applied on the functions $H_{|m|}^{(s)}(\gamma r)$ and $H_{|m|}^{(s)}(\gamma
R)$. For example,
\[\widetilde{E}_{\alpha ,r}^{(1,s)}(r)=-ik\gamma
H_{|m|}^{(s)\prime }(\gamma r).\nonumber
\]

By taking into account (\ref{Gilb2}), for the spectral component of
the Green tensor in (\ref{UDisp}) one finds%
\begin{eqnarray}
&&
G_{jl}^{\rm{(c)}}(\mathbf{r},\mathbf{r};i\xi )=\frac{c^2}{4\pi }%
\sum_{\alpha }\frac{1}{\gamma \omega }\sum_{s=1,2}\left\{ \frac{%
J_{|m|}(\gamma R)}{H_{|m|}^{(s)}(\gamma R)}\right.
\nonumber \\
&&~~~\times
\left[ \frac{E_{\alpha ,j}^{(1,s)}(r)\widetilde{E}_{\alpha
,l}^{(1,s)}(r)}{\omega -i\xi }+\frac{E_{\alpha ,l}^{(1,s)}(r)\widetilde{E}%
_{\alpha ,j}^{(1,s)}(r)}{\omega +i\xi }\right]  \label{Gilb3} \\
&& \left. +\frac{J_{|m|}^{\prime }(\gamma R)}{H_{|m|}^{(s)\prime }(\gamma R)}%
\left[ \frac{E_{\alpha ,j}^{(2,s)}(r)\widetilde{E}_{\alpha ,l}^{(2,s)}(r)}{%
\omega -i\xi }+\frac{E_{\alpha ,l}^{(2,s)}(r)\widetilde{E}_{\alpha
,j}^{(2,s)}(r)}{\omega +i\xi }\right] \right\} .\nonumber
\end{eqnarray}%
Using the expressions for $E_{\alpha ,j}^{(\lambda ,s)}(r)$, for the
off-diagonal components we have
\[
 G_{jl}^{\rm{(c)}}(\mathbf{r},\mathbf{r}%
;i\xi )=-G_{lj}^{\rm{(c)}}(\mathbf{r},\mathbf{r};i\xi ).
\nonumber \]
\noindent
Since the tensor
$\alpha _{jl}(i\xi )$ is symmetric, we conclude that the off-diagonal
components do not contribute to the Casimir-Polder interaction energy.

For the diagonal components, in (\ref{Gilb3}) we rotate the integration
contour in the complex $\gamma $-plane by the angle $\pi /2$ for the $s=1$
term and by the angle $-\pi /2$ for the $s=2$ term. The integrand has poles
at $\gamma =\pm i\sqrt{k^{2}+\xi ^{2}/c^{2}}$ and we assume that they are
avoided by small semicircles $C_{\rho }^{(\pm )}$ of radius $\rho $ in the
right half-plane. As a result the integrals over $\gamma $ for the $s=1$ and
$s=2$ terms are presented as the sum of integrals over the interval $(0,\pm i%
\sqrt{k^{2}+\xi ^{2}/c^{2}}\mp i\rho )$, over $C_{\rho }^{(\pm )}$ and over
the interval $(\pm i\sqrt{k^{2}+\xi ^{2}/c^{2}}\pm i\rho ,\infty )$. It can
be seen that the integrals over the two intervals are cancelled for the $s=1$
and $s=2$ terms. Evaluating the integrals along $C_{\rho }^{(\pm )}$, for
the diagonal components of the Green tensor we find the following expressions%
\begin{eqnarray}
&&
 G_{rr}^{\rm{(c)}}(\mathbf{r},\mathbf{r};i\xi ) =\frac{4}{\pi }%
\sum_{m=0}^{\infty }{\vphantom{\sum}}^{\prime}\int_{0}^{\infty }dk\left[ \frac{\xi ^{2}%
}{c^{2}}\frac{I_{m}^{\prime }(\beta R)}{K_{m}^{\prime }(\beta R)}\frac{m^2
K_{m}^{2}(\beta r)}{\beta^2r^2} \right.
\nonumber \\
&&~~~~~~
 \left. -k^{2}\frac{I_{m}(\beta R)}{K_{m}(\beta R)}%
K_{m}^{\prime 2}(\beta r)\right] ,  \nonumber \\
&&
 G_{\phi \phi }^{\rm{(c)}}(\mathbf{r},\mathbf{r};i\xi )=\frac{4}{\pi }%
\sum_{m=0}^{\infty }{\vphantom{\sum}}^{\prime}\int_{0}^{\infty }dk\left[ \frac{\xi ^{2}%
}{c^{2}}\frac{I_{m}^{\prime }(\beta R)}{K_{m}^{\prime }(\beta R)}%
\,K_{m}^{\prime 2}(\beta r) \right.
\nonumber \\
&&~~~~~~
\left. -k^{2}\frac{I_{m}(\beta R)}{K_{m}(\beta R)}\,\frac{%
m^2K_{m}^{2}(\beta r)}{\beta^2r^2}\right] ,  \label{GilDiag}\\
&&
G_{zz}^{\rm{(c)}}(\mathbf{r},\mathbf{r};i\xi )=-\frac{4}{\pi }%
\sum_{m=0}^{\infty }{\vphantom{\sum}}^{\prime}\int_{0}^{\infty }dk\,\beta ^{2}\frac{%
I_{m}(\beta R)}{K_{m}(\beta R)}\,K_{m}^{2}(\beta r),\nonumber
\end{eqnarray}
\noindent
where $I_{m}(z)$ and $K_{m}(z)$ are the modified Bessel functions and we
introduced the notation $\beta =\sqrt{k^{2}+\xi ^{2}/c^{2}}$. The prime near
the summation sign means that the term with $m=0$ is multiplied by 1/2.

\section{Green tensor for an interior of ideal metal cylindrical shell}

In this section we consider the region inside a cylindrical shell. The
corresponding eigenfunctions have the form (\ref{Ealf}), where the functions
$\mathbf{E}_{\alpha }^{(\lambda )}(r)$ are given by expressions (\ref{Eirn})
with the replacement $g_{\lambda ,\alpha }(\gamma r)\rightarrow
J_{|m|}(\gamma r)$ and for the normalization coefficient one has
\begin{equation}
\beta _{\alpha ,\lambda }^{2}=\frac{c^{2}T_{|m|}(\gamma R)}{\pi R\omega
\gamma },\;T_{\nu }(x)=\frac{x}{J_{\nu }^{^{\prime }2}(x)+(1-\nu
^{2}/x^{2})J_{\nu }^{2}(x)}.  \label{betalf}
\end{equation}%
{}From the boundary conditions on the cylindrical boundary with radius $R$, we
can see that the eigenvalues for the quantum number $\gamma $ are the roots
of the equation $J_{|m|}(\gamma R)=0$ for TM waves and the roots of the
equation $J_{|m|}^{\prime }(\gamma R)=0$ for TE waves. The corresponding
eigenvalues $\gamma R$ we denote $\gamma R=j_{m,n}^{(\lambda )}$, $%
n=1,2,\ldots $, where, as before, $\lambda =1,2$ correspond to the TM and TE
waves. The mode sum for the Green tensor is given by (\ref{GilMod2}), where
now the summation over the collective index $\alpha $ means the summation
over $m$ and $n$, and the integration is performed
over $k$. For the spectral component of
the Green tensor one finds%
\begin{eqnarray}
&&
G_{jl}(\mathbf{r},\mathbf{r};i\xi )=-\frac{c^{2}}{\pi }\,\sum_{\alpha
}\sum_{\lambda =1,2}\frac{T_{|m|}(j_{m,n}^{(\lambda )})}{j_{m,n}^{(\lambda
)}\omega _{m,n}^{(\lambda )}}
\label{GjlInt} \\
&&~~\times
\left[ \frac{E_{\alpha ,j}^{(\lambda
)}(r)E_{\alpha ,l}^{(\lambda )\ast }(r)}{\omega _{m,n}^{(\lambda )}-i\xi }+%
\frac{E_{\alpha ,l}^{(\lambda )}(r)E_{\alpha ,j}^{(\lambda )\ast }(r)}{%
\omega _{m,n}^{(\lambda )}+i\xi }\right] ,
\nonumber
\end{eqnarray}%
where $\omega _{m,n}^{(\lambda )}=c\sqrt{j_{m,n}^{(\lambda )2}/R^{2}+k^{2}}$.

For the summation of the series over $n$ in (\ref{GjlInt}) we apply the
formula \cite{16a}%
\begin{eqnarray}
&&
\sum_{n=1}^{\infty }T_{|m|}(j_{m,n}^{(\lambda )})f(j_{m,n}^{(\lambda )})=%
\frac{1}{2}\int_{0}^{\infty }dx\,f(x)
\label{SumForm}\\
&&~~~
-\frac{\pi i}{2}\sum_{p}%
\left.{\mathrm{Res}}\right|_{z=iy_{p}}f(z)\frac{H_{|m|}^{(1,\lambda )}(z)}{%
J_{|m|}^{(\lambda )}(z)}, \nonumber
\end{eqnarray}%
where
\[
J_{|m|}^{(1)}(z)=J_{|m|}(z), \quad
J_{|m|}^{(2)}(z)=J_{|m|}^{\prime }(z),
\nonumber \]
\noindent
and
\[
H_{|m|}^{(1,1)}(z)=H_{|m|}^{(1)}(z), \quad
H_{|m|}^{(1,2)}(z)=H_{|m|}^{(1)\prime }(z).
\nonumber \]
\noindent
Formula (\ref{SumForm}) is
valid for functions $f(z)$ analytic in the right half-plane $\mathrm{Re\,}%
z>0 $ with poles $\pm iy_{p}$, $y_{p}>0$, $p=1,2,\ldots $, and satisfying
the condition
\[
f(ye^{\pi i/2})=-e^{2|m|\pi i}f(ye^{-\pi i/2})
\nonumber \]
\noindent
 and the condition
 \[
|f(x+iy)|<\varepsilon (x)e^{by}, \quad b<2,
\nonumber \]
\noindent
for $y\rightarrow \infty
$, with $\varepsilon (x)\rightarrow 0$ for $x\rightarrow \infty $. For the
function $f(z)$, corresponding to the components of the Green tensor (\ref%
{GjlInt}), the poles on the imaginary axis correspond to $y_{p}=R\sqrt{%
k^{2}+\xi ^{2}}$ and the above conditions are satisfied if $r<R$. For
example, in the contribution of the TM waves ($\lambda =1$) to the $rr$%
-component of the Green tensor the function $f(z)$ has the form
\[
f(z)=zJ_{|m|}^{\prime 2}(zr/R)/[z^{2}+R^{2}(k^{2}+\xi ^{2})].
\nonumber \]
\noindent
The part of
the Green tensor corresponding to the first term on the right-hand side of (%
\ref{SumForm}) is the Green tensor for the free space. The boundary-induced part
of the Green tensor corresponds to the second term. It can be seen that the
diagonal components of the boundary-induced part in a cylindrical cavity are
obtained from (\ref{GilDiag}) by the replacements $I_{m}\rightleftarrows
K_{m}$.

\section{The Casimir-Polder potential}

We first consider a particle outside of a cylindrical shell. Substituting
(\ref{GilDiag}) into (\ref{UDisp}) we obtain the exact formula for the
Casimir-Polder interaction energy in the general case of an anisotropic
polarizability
\begin{eqnarray}
&&
U(\mathbf{r})=\frac{2\hbar }{\pi
^{2}}\sum_{m=0}^{\infty}{\vphantom{\sum}}^{\prime}
\int_{0}^{\infty }d\xi \,\,\int_{\xi/c}^{\infty }d\beta \frac{\beta }{\sqrt{%
\beta ^{2}-\xi ^{2}/c^{2}}} \nonumber \\
&&~~
\times \left\{ \alpha _{rr}(i\xi )\left[ \frac{\xi ^{2}}{c^{2}}\frac{%
I_{m}^{\prime }(\beta R)}{K_{m}^{\prime }(\beta R)}\frac{m^{2}K_{m}^{2}(%
\beta r)}{\beta ^{2}r^{2}}\right.\right.
\nonumber \\
&&~~~~~~~\left.
-\left( \beta ^{2}-\frac{\xi ^{2}}{c^{2}}\right)
\frac{I_{m}(\beta R)}{K_{m}(\beta R)}K_{m}^{\prime 2}(\beta r)\right]
\nonumber \\
&&~~~~
+\alpha _{\phi \phi }(i\xi )\left[ \frac{\xi ^{2}}{c^{2}}\frac{%
I_{m}^{\prime }(\beta R)}{K_{m}^{\prime }(\beta R)}\,K_{m}^{\prime 2}(\beta
r)\right.
\nonumber \\
&&~~~~~~~\left.
-\left( \beta ^{2}-\frac{\xi ^{2}}{c^{2}}\right) \frac{I_{m}(\beta R)}{%
K_{m}(\beta R)}\,\frac{m^{2}K_{m}^{2}(\beta r)}{\beta ^{2}r^{2}}\right]
 \nonumber \\
&&~~~~ \left.
\vphantom{\left[\frac{I_{m}^{\prime }(\beta R)}{K_{m}^{\prime }(\beta R)}\right]}
-\alpha _{zz}(i\xi )\beta ^{2}\frac{I_{m}(\beta R)}{K_{m}(\beta R)}%
\,K_{m}^{2}(\beta r)\right\} ,\label{eq17a}
\end{eqnarray}%
\noindent where $\alpha _{rr}$, $\alpha _{\phi \phi }$, and $\alpha _{zz}$
are the diagonal components of an atomic (molecular) polarizability.

In the isotropic case the exact Casimir-Polder energy (\ref{eq17a})
simplifies to%
\begin{eqnarray}
&&
 U(r)=\frac{2\hbar }{\pi ^{2}}\,\sum_{m=0}^{\infty}
{\vphantom{\sum}}^{\prime}
\int_{0}^{\infty \!\!}d\xi \,\alpha (i\xi )\int_{\xi /c}^{\infty }
\!\!d\beta \frac{%
\beta }{\sqrt{\beta ^{2}-\xi ^{2}/c^{2}}}  \nonumber \\
\hspace*{-2cm}
&&\times \left\{ \frac{\xi ^{2}}{c^{2}}\frac{I_{m}^{\prime }(\beta R)}{%
K_{m}^{\prime }(\beta R)}G_{m}(\beta r)-\,\frac{I_{m}(\beta R)}{K_{m}(\beta
R)} \right.
\label{UDispExt2} \\
&&~~~~~~~~\left.\times
 \left[ \left(\beta ^{2}-{\xi ^{2}}/{c^{2}}\right)G_{m}(\beta
r)+\beta ^{2}K_{m}^{2}(\beta r)\right]
\vphantom{\frac{\xi^2}{C^2}}
\right\} ,  \nonumber
\end{eqnarray}%
with the notation $G_{m}(z)=K_{m}^{\prime 2}(z)+\left( m/z\right)
^{2}\,K_{m}^{2}(z). $

It can be easily seen that (\ref{eq17a}) coincides with expression (17)
[with notations (27)--(29)] of
\cite{21b}, where the exact Casimir-Polder potential for an atom outside
an ideal metal cylinder was obtained. To see this we use the following
expressions for the components of atomic polarizability
\begin{equation}
\alpha _{ll}(\omega)=2\sum_{j}\frac{E_{ji}|\left\langle j\left\vert \mu
_{l}\right\vert i\right\rangle |^{2}}{E_{ji}^{2}-\hbar ^{2}\omega ^{2}}.
\label{eq18aa}
\end{equation}
\noindent
Here, $|i\rangle $ is the state of an atom with the energy $E_{i}$%
, $E_{ji}\equiv E_{j}-E_{i}$, and {\boldmath$\mu $} is the
atomic dipole moment
operator. Now we substitute (\ref{eq18aa}) into (\ref{eq17a}) and return to
the integration variable $k$:
\begin{eqnarray}
&&
U(\mathbf{r}) =\frac{4\hbar }{\pi ^{2}}\sum_{j}E_{ji}
\sum_{m=0}^{\infty }{\vphantom{\sum}}^{\prime}
\int_{0}^{\infty }d\xi \,\,\int_{0}^{\infty }dk\frac{1}{%
E_{ji}^{2}+\hbar ^{2}\xi ^{2}}  \nonumber \\
&&\!\times \!\left\{ \!|\mu _{r}|^{2}\left[ \frac{\xi ^{2}}{c^{2}}\frac{%
I_{m}^{\prime }(\beta R)}{K_{m}^{\prime }(\beta R)}\frac{m^{2}K_{m}^{2}(%
\beta r)}{\beta ^{2}r^{2}}-k^{2}\frac{I_{m}(\beta R)}{K_{m}(\beta R)}%
K_{m}^{\prime 2}(\beta r)\right] \right.   \nonumber \\
&& +|\mu _{\phi }|^{2}\left[ \frac{\xi ^{2}}{c^{2}}\frac{I_{m}^{\prime
}(\beta R)}{K_{m}^{\prime }(\beta R)}\,K_{m}^{\prime 2}(\beta r)-k^{2}\frac{%
I_{m}(\beta R)}{K_{m}(\beta R)}\,\frac{m^{2}K_{m}^{2}(\beta r)}{\beta
^{2}r^{2}}\right]
\nonumber \\
&&\left.
-|\mu _{z}|^{2}\beta ^{2}\frac{I_{m}(\beta R)}{K_{m}(\beta
R)}\,K_{m}^{2}(\beta r)\right\}. \label{eq18b}
\end{eqnarray}%
\noindent
Next we introduce polar coordinates in the plane $(\xi /c,k)$: $%
\xi /c=\beta \,\cos \theta $, $k=\beta \sin \theta $. This allows us to
rearrange (\ref{eq18b}) into the form
\begin{eqnarray}
&&
U(\mathbf{r})=\frac{4\hbar c}{\pi ^{2}}\sum_{j}E_{ji}
\sum_{m=0}^{\infty }{\vphantom{\sum}}^{\!\!\prime}
\int_{0}^{\infty }\!\!d\beta \,\,\int_{0}^{\pi /2} \!\!\!\frac{%
\beta ^{3}d\theta}{E_{ji}^{2}+\hbar ^{2}c^{2}\beta ^{2}\cos ^{2}\theta }
\nonumber \\
&&\times \left\{ |\mu _{r}|^{2}\left[ \cos ^{2}\theta \frac{I_{m}^{\prime
}(\beta R)}{K_{m}^{\prime }(\beta R)}\frac{m^{2}K_{m}^{2}(\beta r)}{\beta
^{2}r^{2}}-\sin ^{2}\theta\right.\right.
\label{eq18c} \\
&&\left.
\times \frac{I_{m}(\beta R)}{K_{m}(\beta R)}%
K_{m}^{\prime 2}(\beta r)\right]
+|\mu _{\phi }|^{2}\left[ \cos ^{2}\theta \frac{I_{m}^{\prime
}(\beta R)}{K_{m}^{\prime }(\beta R)}\,K_{m}^{\prime 2}(\beta r)
\right.
\nonumber \\
&&\left.
-\sin
^{2}\theta \frac{I_{m}(\beta R)}{K_{m}(\beta R)}\,\frac{m^{2}K_{m}^{2}(\beta
r)}{\beta ^{2}r^{2}}\right]
\left.\!\!
-|\mu _{z}|^{2}\frac{I_{m}(\beta R)}{K_{m}(\beta
R)}\,K_{m}^{2}(\beta r)\!\right\}\! . \nonumber
\end{eqnarray}%
\noindent
Evaluating the integrals over $\theta $, we obtain from
(\ref{eq18c}) the following result:
\begin{eqnarray}
&&
U(\mathbf{r})=-\frac{2\hbar }{\pi c}\sum_{j}
\sum_{m=0}^{\infty }{\vphantom{\sum}}^{\prime}
\int_{0}^{\infty }d\beta \,\beta \,\,
\nonumber  \\
&&\times
\left\{ |\mu _{z}|^{2}%
\frac{c^{2}\beta ^{2}}{\sqrt{E_{ji}^{2}+\hbar ^{2}c^{2}\beta ^{2}}}\frac{%
I_{m}(\beta R)}{K_{m}(\beta R)}\,K_{m}^{2}(\beta r)\right.
\label{eq18d}  \\
&&+|\mu _{r}|^{2}\!\left[ \!\left( \frac{E_{ji}}{\sqrt{E_{ji}^{2}+\hbar
^{2}c^{2}\beta ^{2}}}-1\right)\! E_{ji}\frac{I_{m}^{\prime }(\beta R)}{%
K_{m}^{\prime }(\beta R)}\frac{m^{2}K_{m}^{2}(\beta r)}{\beta ^{2}r^{2}}%
\right.
\nonumber \\
&&~
+\left.\left( \sqrt{E_{ji}^{2}+\hbar ^{2}c^{2}\beta ^{2}}-E_{ji}\right) \frac{%
I_{m}(\beta R)}{K_{m}(\beta R)}K_{m}^{\prime 2}(\beta r)
\vphantom{\left( \frac{E_{ji}}{\sqrt{E_{ji}^{2}+\hbar^{2}c^{2}\beta ^{2}}}\right)}
\right]   \nonumber
\\
&& +|\mu _{\phi }|^{2}\left[ \left( \frac{E_{ji}}{\sqrt{%
E_{ji}^{2}+\hbar ^{2}c^{2}\beta ^{2}}}-1\right) E_{ji}\frac{I_{m}^{\prime
}(\beta R)}{K_{m}^{\prime }(\beta R)}\,K_{m}^{\prime 2}(\beta r)
\right.
\nonumber
\end{eqnarray}
\begin{eqnarray}
&&\left.\left.
+\left(
\sqrt{E_{ji}^{2}+\hbar ^{2}c^{2}\beta ^{2}}-E_{ji}\right) \frac{I_{m}(\beta
R)}{K_{m}(\beta R)}\,\frac{m^{2}K_{m}^{2}(\beta r)}{\beta ^{2}r^{2}}
\vphantom{\left( \frac{E_{ji}}{\sqrt{E_{ji}^{2}+\hbar^{2}c^{2}\beta ^{2}}}\right)}
\right]\!
\right\}\! .\nonumber
\end{eqnarray}%
\noindent The latter coincides with the result given by equations (17),
(27)-(29) of \cite{21b} (note that in \cite{21b} the
International system of units
is used).

For an anisotropic particle inside a cylindrical shell the Casimir-Polder
potential is obtained by the substitution of the Green tensor (\ref{GjlInt})
in (\ref{UDisp}), which results in
\begin{eqnarray}
&&
U(\mathbf{r})=\frac{2\hbar }{\pi ^{2}}
\sum_{m=0}^{\infty}{\vphantom{\sum}}^{\prime}
\int_{0}^{\infty }d\xi \,\,\int_{\xi/c}^{\infty }d\beta \frac{\beta }{\sqrt{%
\beta ^{2}-\xi ^{2}/c^{2}}}  \nonumber \\
&&\times \left\{ \alpha _{rr}(i\xi )\left[ \frac{\xi ^{2}}{c^{2}}\frac{%
K_{m}^{\prime }(\beta R)}{I_{m}^{\prime }(\beta R)}\frac{m^{2}I_{m}^{2}(%
\beta r)}{\beta ^{2}r^{2}}\right.\right.
\nonumber \\
&&~~~~~~~~\left.
-\left( \beta ^{2}-\frac{\xi ^{2}}{c^{2}}\right)
\frac{K_{m}(\beta R)}{I_{m}(\beta R)}I_{m}^{\prime 2}(\beta r)\right]
\nonumber \\
&& +\alpha _{\phi \phi }(i\xi )\left[ \frac{\xi ^{2}}{c^{2}}\frac{%
K_{m}^{\prime }(\beta R)}{I_{m}^{\prime }(\beta R)}\,I_{m}^{\prime 2}(\beta
r)\right.
\nonumber \\
&&~~~~~~~~\left.
-\left( \beta ^{2}-\frac{\xi ^{2}}{c^{2}}\right) \frac{K_{m}(\beta R)}{%
I_{m}(\beta R)}\,\frac{m^{2}I_{m}^{2}(\beta r)}{\beta ^{2}r^{2}}\right]
\nonumber \\
&&\left.
-\alpha _{zz}(i\xi )\beta ^{2}\frac{K_{m}(\beta R)}{I_{m}(\beta R)}%
\,I_{m}^{2}(\beta r)\right\} .  \label{eq18e}
\end{eqnarray}%
\noindent
The Casimir-Polder energy for an isotropic particle inside a
cylindrical shell is given by the expression
\begin{eqnarray}
&&U(r)=\frac{2\hbar }{\pi ^{2}}\,
\sum_{m=0}^{\infty}{\vphantom{\sum}}^{\prime}
\int_{0}^{\infty }d\xi \,\alpha (i\xi )\int_{\xi /c}^{\infty }d\beta \frac{%
\beta }{\sqrt{\beta ^{2}-\xi ^{2}/c^{2}}}
\nonumber \\
&&~~\times \left\{ \frac{\xi ^{2}}{c^{2}}\frac{K_{m}^{\prime }(\beta R)}{%
I_{m}^{\prime }(\beta R)}F_{m}(\beta r)-\,\frac{K_{m}(\beta R)}{I_{m}(\beta
R)}\right.
\label{UDispInt2} \\
&&~~~~
\left.\times
\left[ \left( \beta ^{2}-\xi ^{2}/c^{2}\right) F_{m}(\beta
r)+\beta ^{2}I_{m}^{2}(\beta r)\right]
\vphantom{\frac{\xi ^{2}}{c^{2}}}
\right\} ,  \nonumber
\end{eqnarray}%
where $F_{m}(z)=I_{m}^{\prime 2}(z)+\left( m/z\right) ^{2}\,I_{m}^{2}(z)$.
We emphasize that the results (\ref{eq18e}) and (\ref{UDispInt2})
cannot be obtained from the formalism \cite{26a} developed for an
atom inside a hollow dielectric cylinder of finite permittivity.

\section{The role of dynamic effects and the comparison with the proximity
force approximation}

The following analysis is given for the
expression (\ref{UDispExt2}), i.e.,
for an isotropic polarizable particle spaced outside  a cylindrical shell.

Let us denote by $a$ the separation distance between a particle and a surface
of the cylinder, so that $a=r-R$. At small distances, $a/R\ll 1$, the
dominant contribution to Eq.~(\ref{UDispExt2}) comes from large values of $m$
and we can use the uniform asymptotic expansions for the modified Bessel
functions for large values of $m$. After calculations, in two leading orders
in $a/R$, we find
\begin{eqnarray}
&& U(a) \approx -\frac{\hbar c}{16\pi a^{4}}\,\int_{0}^{\infty }\!\!d\zeta
\,\alpha (i\omega _{c}\zeta )
\left\{ \vphantom{\frac{a}{2R}}(2+2\zeta
+\zeta ^{2})e^{-\zeta }\right.  \label{Usmall}\\
&&\left. -\frac{a}{2R}\,\left[ (2+2\zeta +\zeta ^{2})e^{-\zeta }+2\zeta
^{2}\Gamma (0,\zeta )-\zeta ^{4}\Gamma (-2,\zeta )\right] \right\} ,
\nonumber
\end{eqnarray}%
where $\omega _{c}=c/(2a)$, $\zeta =\xi /\omega _{c}$, and $\Gamma (b,\zeta
) $ is the incomplete gamma function. The term of
zeroth order in $a/R$  in (\ref{Usmall})
coincides with the Casimir-Polder potential for an
atom near an ideal metal plate (see equation~(16.27) in \cite{17}).

The oscillator model for the dynamic polarizability of a particle%
\begin{equation}
\alpha (i\xi )=\sum_{j}\frac{g_{j}}{\omega _{j}^{2}+\xi ^{2}},  \label{alfa}
\end{equation}%
where $g_{j}$ are the oscillator strengths and $\omega _{j}$ are the
eigenfrequencies, works well over a wide range of separations \cite{7,8}.
Substituting (\ref{alfa}) into (\ref{UDispExt2}) and evaluating the
integrals over $\xi $ explicitly, one finds%
\begin{eqnarray}
&&
U(r) =\frac{\hbar c}{\pi }\sum_{j}\frac{g_{j}}{\omega _{j}^{2}}%
\sum_{m=0}^{\infty }{\vphantom{\sum}}^{\!\prime}
\,\int_{0}^{\infty \!\!}d\beta \,\beta
^{3}
\left\{ \frac{I_{m}^{\prime }(\beta R)}{K_{m}^{\prime }(\beta R)}%
\right.
\label{UDispExt3} \\
&&\times
\frac{G_{m}(\beta r)}{s_j(\beta)[s_j(\beta)+1]}
\left. -\frac{I_{m}(\beta R)}{K_{m}(\beta R)}\left[ \,\frac{%
K_{m}^{2}(\beta r)}{s_j(\beta)}+\frac{G_{m}(\beta r)}{s_j(\beta)+1}\right]
\right\} ,\nonumber
\end{eqnarray}%
\noindent where $s_j(\beta)\equiv\sqrt{c^2\beta^2/\omega_j^2+1}$. The main
contribution to the $\beta $-integral in (\ref{UDispExt3}) comes from $%
\beta \lesssim 1/a$. If $a\gg c/\omega _{j}$, the ratio $c\beta /\omega _{j}$
is small and from (\ref{UDispExt3}) we obtain
\begin{eqnarray}
&& U^{\rm (stat)}(r)=\hbar c\frac{\alpha _{0}}{2\pi }
\sum_{m=0}^{\infty }{\vphantom{\sum}}^{\prime}
\int_{0}^{\infty }d\beta \,\beta ^{3}\left\{ \frac{%
I_{m}^{\prime }(\beta R)}{K_{m}^{\prime }(\beta R)}G_{m}(\beta r) \right.
\nonumber \\
&&~~~~\left. -\frac{I_{m}(\beta R)}{K_{m}(\beta R)}\left[ G_{m}(\beta
r)+2K_{m}^{2}(\beta r)\right] \right\} ,  \label{Uext}
\end{eqnarray}%
where $\alpha _{0}=\sum_{j}g_{j}/\omega _{j}^{2}$ is the static
polarizability of a particle.
In the static limit, assuming $a/R\ll 1$, from (\ref{Usmall})
 we find%
\begin{equation}
U^{\rm{(stat)}}(a)\approx U_{0}(a)\left( 1-\frac{26}{45}\frac{a}{R}\right)
,  \label{UsmallSt}
\end{equation}%
where $U_{0}(a)=-3\hbar c\alpha _{0}/(8\pi a^{4})$ is the famous
Casimir-Polder result for the interaction energy between a particle and
an ideal metal plane \cite{1}.

\begin{figure}[t]
\vspace*{0.5cm}
\resizebox{0.45\textwidth}{!}{%
\includegraphics{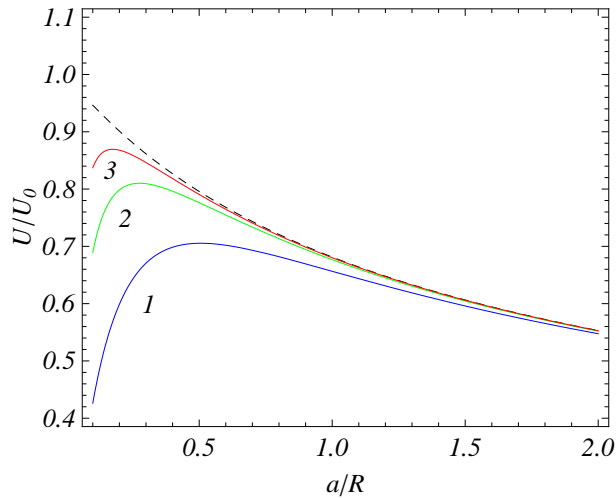}
} \caption{\label{fig1} The normalized exact Casimir-Polder
potential for a particle outside a cylindrical
shell as a function of $a/R$ for
$R/\lambda _{A}=1,\>2.5$ and 5
(solid lines 1, 2, and 3, respectively).
The dashed line represents the Casimir-Polder potential in the
static limit.}
\end{figure}
\begin{figure}[b]
\vspace*{0.5cm}
\resizebox{0.45\textwidth}{!}{%
\includegraphics{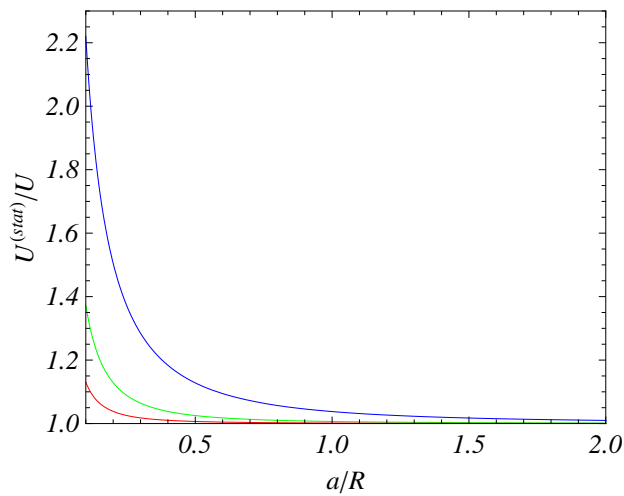}}
 \caption{\label{fig2} The ratio of exact Casimir-Polder
potentials calculated using static and dynamic
polarizabilities of a particle
as a function of $a/R$ for the same values of
$R/\lambda _{A}$, as in Fig.~\ref{fig1}, increasing from top to
bottom lines.}
\end{figure}
In Fig.~\ref{fig1} we plot the exact Casimir-Polder energy (\ref{UDispExt3})
for a particle outside a shell
normalized for $U_{0}$ as a function of $a/R$ using the single-oscillator
model [(\ref{alfa}) with $\omega _{1}=\omega _{A}$]. Different lines are
plotted for $R/\lambda _{A}=1,\>2.5$, and $5$ (lines 1, 2, and 3
respectively), where $\lambda _{A}=2\pi c/\omega _{A}$. The dashed line
corresponds to the static limit (\ref{Uext}). For instance, for an
atom of metastable He$^{\ast }$ one has \cite{8} $\omega _{A}=1.18$ eV and
the cylinder radius corresponding to $R/\lambda _{A}=1$ is equal to $%
0.95\,\mu$m. As can be seen in Fig.~\ref{fig1}, with increasing $R/\lambda
_{A}$ the Casimir-Polder energy approaches to the static limit. Note that
although lines 1, 2, and 3 around $a/R=0$ demonstrate the same distance
dependence (the inverse third power of separation) as does the
nonrelativistic limit (i.e., the van der Waals force), they are quite
different in nature. Particularly, the van der Waals force does not depend
on $c$ whereas the potential (\ref{UDispExt3}) in the limit $a/R\to 0$ does. In
fact, the nonrelativistic limit cannot be achieved for an ideal metal
cylinder. Figure~\ref{fig2} shows the ratio $U^{\rm{(stat)}}/U$ as a
function of $a/R$ (the upper, middle and lower lines correspond to $%
R/\lambda _{A}=1,\>2.5$, and $5$, respectively).

\begin{figure}[t]
\vspace*{0.5cm}
\resizebox{0.45\textwidth}{!}{%
\includegraphics{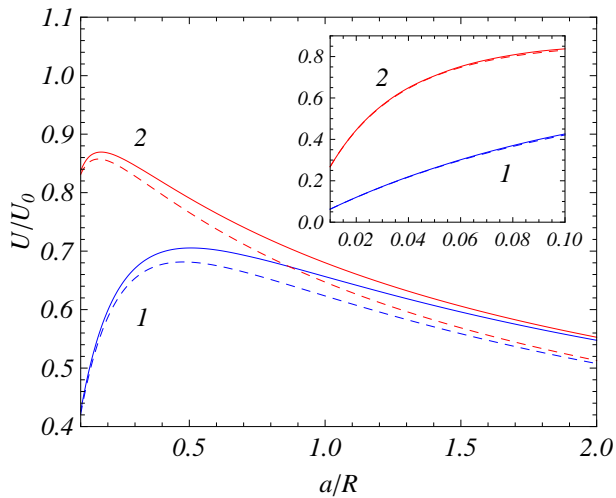}
} \caption{\label{fig3} Comparison between the exact and PFA
Casimir-Polder potentials for particle-cylinder interaction (the solid
and dashed lines, respectively) for $R/\lambda _{A}=1$ and 5
(lines labeled 1 and 2) in the separation region $0.1 \leq
a/R \leq 2$. The inset shows the region from $a/R=0.01$ to
$a/R=0.1$.}
\end{figure}

Now we turn to the comparison between the exact Casimir-Polder energy with
that obtained using PFA. The latter is given by the expression%
\begin{eqnarray}
&& U_{\rm{PFA}}(a)=-\frac{\hbar c}{16\pi a^{4}}\sqrt{\frac{R}{R+a}}%
\int_{0}^{\infty }d\zeta \,\alpha (i\omega _{c}\zeta )e^{-\zeta}  \nonumber
\\
&&~~~~~~\times \left[ 2+2\zeta +\zeta ^{2}-\frac{a(1+\zeta )}{2(R+a)}\right]
\label{UPFA}
\end{eqnarray}%
obtained from corresponding equation in  \cite{14} by putting the
reflection coefficients equal to $\pm 1$, as it holds for ideal metal. In
the static limit, (\ref{UPFA}) leads to%
\begin{equation}
U_{\rm{PFA}}(a)=U_{0}(a)\sqrt{\frac{R}{R+a}}\left( 1-\frac{1}{6}\frac{a}{%
R+a}\right) ,  \label{UPFAstat}
\end{equation}%
and for $a/R\ll 1$ one has
\begin{equation}
U_{\rm{PFA}}(a)\approx U_{0}(a)\left( 1-\frac{2a}{3R}\right),
\label{eq27a}
\end{equation}
\noindent
to be compared with the exact result (\ref{UsmallSt}) obtained in
this limiting case. In Fig.~\ref{fig3} we plot the ratio $U/U_{0}$ evaluated
in the single-oscillator model by using exact
expression (\ref{UDispExt3}) (solid lines)
and by the PFA, expression (\ref{UPFA}) (dashed lines). The lines 1 and 2
correspond to $R/\lambda _{A}=1$ and $5$, respectively. As can be seen in
Fig.~\ref{fig3}, the PFA results are in very good agreement with the exact
results. For example, at $a/R=0.1$ the relative deviation between the two
results is equal to 1.3\% and 0.9\% for lines 1 and 2, respectively. The
same lines within the region $0.01\leqslant a/R\leqslant 0.1$ are shown on
an inset to Fig.~\ref{fig3}. At $a/R=0.02$ the relative deviation between
the exact and the PFA results does not exceed 0.4\%.

\section{Microparticle in an interior of cylindrical shell}

Here, we present a few computational results for a microparticle
inside an ideal metal cylindrical shell.
Substituting (27) into (25) and integrating over $\xi $, we find
the expression%
\begin{eqnarray}
&&
U(r) =\frac{\hbar c}{\pi }\sum_{j}\frac{g_{j}}{\omega _{j}^{2}}%
\sum_{m=0}^{\infty }{\vphantom{\sum}}^{\!\prime }\,\int_{0}^{\infty }
\!\!d\beta
\,\beta ^{3}
\left\{ \frac{K_{m}^{\prime }(\beta R)}{I_{m}^{\prime }(\beta R)}%
\right.
\label{UDispInt3} \\
&&\times
\frac{F_{m}(\beta r)}{s_{j}(\beta )[s_{j}(\beta )+1]}
\left. -\frac{K_{m}(\beta R)}{I_{m}(\beta R)}\left[ \,\frac{%
I_{m}^{2}(\beta r)}{s_{j}(\beta )}+\frac{F_{m}(\beta r)}{s_{j}(\beta )+1}%
\right] \right\} . \nonumber
\end{eqnarray}
\noindent
In the static limit, similar to (29), one obtains
\begin{eqnarray}
&&
U^{\rm (stat)}(r) =\hbar c\frac{\alpha_0}{2\pi }%
\sum_{m=0}^{\infty }{\vphantom{\sum}}^{\!\prime }\,\int_{0}^{\infty }
\!\!d\beta
\,\beta ^{3}
\left\{ \frac{K_{m}^{\prime }(\beta R)}{I_{m}^{\prime }(\beta R)}%
F_{m}(\beta r)\right.
\nonumber \\
&&
\left. -\frac{K_{m}(\beta R)}{I_{m}(\beta R)}\left[F_{m}(\beta r)
+2I_{m}^{2}(\beta r)%
\right] \right\} .\label{eq35}
\end{eqnarray}

\begin{figure}[b]
\vspace*{0.5cm}
\resizebox{0.45\textwidth}{!}{%
\includegraphics{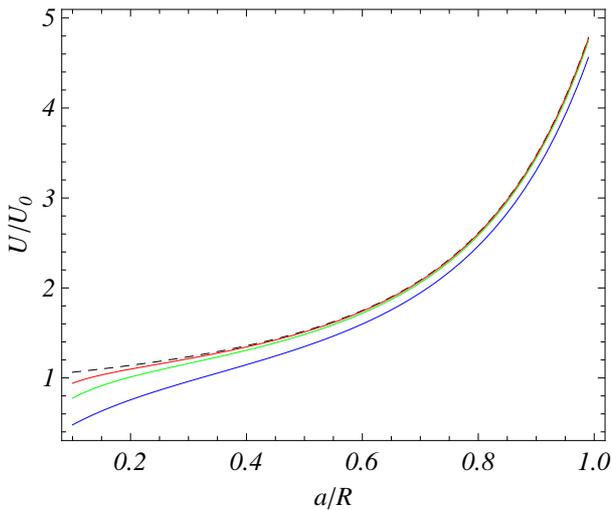}
} \caption{\label{fig4} The normalized exact Casimir-Polder
potential for a particle inside a cylindrical shell
as a function of $a/R$ for
$R/\lambda _{A}=1,\>2.5$ and 5 increasing from bottom to
top solid lines.
The dashed line represents the Casimir-Polder potential in the
static limit.
}
\end{figure}

The computational results are presented in Fig.~\ref{fig4}, where
the exact Casimir-Polder energy (\ref{UDispInt3}) normalized for
$U_0$ is plotted as a function of $a/R$ (note that in the interior
of a cylindrical shell $a=R-r$ holds).
In so doing the single-oscillator model was again used.
Similar to Fig.~\ref{fig1}, the solid lines from the bottom
to the top one correspond to $R/\lambda_A=1,\>2.5$, and 5,
respectively.
The dashed line represents the Casimir-Polder potential in the
static limit (\ref{eq35}). As is seen in Fig.~\ref{fig4},
with increasing $R/\lambda_A$ the Casimir-Polder energy
for a particle inside a cylindrical shell approaches  the
static limit. This is in analogy to a particle outside
an ideal metal cylindrical shell.

\section{Conclusions and Discussion}

In the foregoing we have derived the exact Casimir-Polder potential for a
polarizable particle placed in an exterior or in an interior of an ideal metal
cylindrical shell. Derivation was performed by the Green tensor method. The
obtained results for a particle outside the cylindrical shell were compared with
obtained by another method and found to be in agreement. Computations were
made for particles described by the single-oscillator model. It was shown that
at particle-cylinder separations below 0.1$R$ errors introduced by the use of
the PFA do not exceed 1.3\%. This justifies the application of the PFA for
the comparison of the experimental data with theory.
We have also investigated the role of dynamic polarizability
of the particle.
According to our computational results, the static limit is approached
for large $a/R$ or $R/\lambda_A$.
Replacing ideal metal
boundary conditions used in expressions (\ref{Ealf}) and (\ref{Eirn}) by, for
instance, standard electrodynamic continuity conditions or boundary
conditions of the Dirac model \cite{18,25a}, it is possible to find
the respective exact solutions for a particle interacting with a dielectric
cylindrical shell or with a single-walled carbon nanotube.

The obtained
results can be used in the theory of topological defects, e.g., when
there is a cosmic string along the axis of the cylinder. For a cosmic string
with the angle deficit $2\pi -\phi _0$ ($0\leqslant \phi \leqslant \phi _0$%
), the formulae for the Casimir-Polder potential are obtained from (\ref%
{UDispExt2}) and (\ref{UDispInt2}) by the replacement $m \rightarrow qm$ and
adding the factor $q$, where $q=2\pi /\phi _0$ (for the corresponding
eigenfunctions see \cite{19}). Note that the conducting cylindrical
surface can be considered as a simple model of superconducting string core.
Superconducting strings are predicted in a wide class of field theories and
they are sources of a number of interesting astrophysical effects such as
generation of synchrotron radiation, cosmic rays, and relativistic jets.

\section*{Acknowledgments}

The authors thank CNPq (Brazil) for partial financial support.
G.L.K.\ was also partially
supported by the Grant of the Russian Ministry of Education P--184.
 A.A.S.\ was partially supported by the Armenian Ministry of Education and Science,
Grant No. 119.


\end{document}